\documentclass[12pt]{article}
\def\1{\'{\i}}

\usepackage[cm]{fullpage}
\usepackage{amssymb}
\usepackage{authblk}

\title{Approximate Kerr-Newman-like Metric with Quadrupole}
\author[1,*]{\rm Francisco Frutos-Alfaro}
\author[2]{Paulo Montero-Camacho}
\date{\today}

\affil[1]{\small{School of Physics and Space Research Center of University of 
Costa Rica}}
\affil[2]{\small{Center for Cosmology and AstroParticle Physics, The Ohio State 
University}}
\affil[*]{E-mail: \textit {frutos@fisica.ucr.ac.cr}}

\begin{document}
\maketitle

\abstract{
The Kerr metric is known to present issues when trying to find an interior 
solution. In this work we continue in our efforts to construct a more realistic 
exterior metric for astrophysical objects. A new approximate metric 
representing the spacetime of a charged, rotating and slightly-deformed body 
is obtained by perturbing the Kerr-Newman metric to include the mass-quadrupole 
and quadrupole-quadrupole orders. It has a simple form, because is 
Kerr-Newman-like. Its post-linear form without charge coincides with 
post-linear quadrupole-quadrupole metrics already found.}

\section{Introduction}
\label{sec:Intro}

\noindent
Since Kerr proposed his metric in 1963 \cite{Kerr} multiple efforts have been 
directed to finding an interior solution of his spacetime 
\cite{Cuchi, Haggag1, Haggag2, Haggag3, Hernandez, Krasinski, Krisch, Ramadan} 
or in generalizing the Kerr metric to a metric that allows a physical interior 
matching \cite{HT,Thorne, QM,Quevedo}. Nonetheless, no physical interior 
solution exists. See \cite{Boshkayev} for a relatively recent prespective in 
the issues present in the Kerr metric that complicate the interior matching.

\noindent 
Furthermore, from some of the attempts of constructing an interior metric of 
the Kerr metric we see a trend for preference of an oblate spheroid instead of 
a sphere \cite{Viaggiu,Drake}. This is a mathematical motivation indicating 
the value of exploring metrics of deformed objects. However, the physical 
motivation is much more simple, real astrophysical objects are not perfect 
spheres, hence allowing for small deformations in a rotating object is 
meaningful. 

\noindent
Moreover, the interest in spacetimes capable of describing charged 
distributions has always been high. In the early 1916's  Reissner and 
Nordstr\"om \cite{Nordstrom, Reissner} found their metric, which described a 
static spherically symmetric charge distribution. Even though this metric does 
not handle rotation it has been used, for example, to do black hole lensing 
\cite{Eiroa} and study Hawking radiation from a Reissner-Nordstr\"om black hole 
\cite{Zhang}. It is important to highlight that this kind of studies would 
benefit from a metric capable of describing charged objects but including 
rotation, and with the capabilities of allowing deformed objects, this last 
property rules out the Kerr-Newman metric \cite{Newman}. 

\noindent
In this article we continue our efforts in constructing a more realistic metric 
capable of representing a real astrophysical object. Here, we use a 
perturbative method which utilizes the Lewis metric \cite{Carmeli} in order to 
find spacetimes with quadrupole moment while using the Kerr spacetime as a seed 
metric. Basically, our technique consists in cleverly changing the potentials 
of the Lewis metric while maintaining the cross term (rotational term). 
We have already applied this techinque and obtained other approximate metrics 
\cite{Frutos1,Frutos2,Montero}. In comparison to our previous efforts this work 
includes the addition of charge. Hence, our new metric is capable of 
representing a charged, rotating and  slightly deformed massive object.

\noindent
The usual question when computing new solutions of the Einstein Field Equations 
(EFE) is how to prove that a given metric would have physical 
meaning\footnote{Here physical meaning stands for having an interior metric.}. 
In order to confirm the physical legitimacy of a given metric one can expand 
it to its post-linear from and compare the result with the post-linear version 
of the Hartle-Thorne (HT) metric \cite{HT, Quevedo}. Note that it is possible 
to find an inner solution to the HT metric \cite{Boshkayev}, and hence if a 
metric has a similar form to the HT metric then an inner solution should 
exists. See \cite{Frutos3,Frutos4} for a more detailed discussion and some 
examples. 

\noindent
The rest of this paper is organized as follows: We start our perturbation 
method of the Kerr metric with the help of the Lewis potentials in section 
\ref{sec:PKM}. In section \ref{sec:super}, we obtained a new metric by means 
of our perturbative technique, this metric has rotation, quadrupole moments and 
charge. We checked that the metric is a solution of the EFE using a REDUCE 
program \cite{Hearn}, this program is available upon request. We state our 
conclusions in section \ref{sec:conclusion}.

\section{The Perturbing Method for the Kerr Metric}
\label{sec:PKM}

\noindent
Here and in the following sections we will use the method developed by 
Frutos {et al.} in \cite{Frutos1,Frutos2,Montero} to obtain a Kerr-Newman-like 
metric {\it i.e.} a spacetime capable of describing a slightly deformed 
rotating charged mass. We start by using the connection between the Lewis 
metric and the Kerr metric to obtain the Lewis potentials associated to our 
new metric. These potentials are later used in the perturbation method. 

\noindent 
First of all, we start with the Lewis metric, which is given by \cite{Carmeli}

\begin{equation}
\label{lewis} 
{d}{s}^2 = - V d t^2 + 2 W d t d \phi + X d \rho^2 + Y d z^2 + Z d \phi^2 ,
\end{equation} 

\noindent
where the chosen canonical co\-or\-di\-na\-tes are $ x^{1} = \rho $ and 
$ x^{2} = z $. The potentials $ V, \, W, \, Z $, $ X = {\rm e}^{\mu} $ and 
$ Y = {\rm e}^{\nu} $ are functions of $ \rho $ and $ z $ with 
$ \rho^2 = V Z + W^2 $. 

\noindent
From \cite{Carmeli} the transformation that leads to the Kerr metric is

\begin{equation}
\label{chandra} 
\rho = \sqrt{\Delta} \sin{\theta} \qquad {\rm and} 
\qquad z = (r - M) \cos{\theta} ,
\end{equation} 

\noindent
where $ \Delta = r^2 - 2 M r + a^2 + e^2 $, $a$ is the rotational parameter and 
$ e $ is the electric charge. 
 
\noindent
Now, the Lewis potentials are chosen as follows 

\begin{eqnarray}
\label{potential}
V & = & V_K \, {\rm e}^{- 2 \psi} 
= \frac{1}{{\tilde{\rho}}^2} [\Delta - a^2 \sin^2{\theta}] \, {\rm e}^{- 2 \psi} 
\nonumber \\
W & = & - \frac{2 J r}{{\tilde{\rho}}^2} \sin^2{\theta} \nonumber \\
X & = & X_K {\rm e}^{2 \chi} = {\tilde{\rho}}^2 \frac{{\rm e}^{2 \chi}}{\Delta} \\
Y & = & Y_K {\rm e}^{2 \chi} = {\tilde{\rho}}^2 \, {\rm e}^{2 \chi} \nonumber \\
Z & = & Z_K {\rm e}^{2 \psi} = \frac{\sin^2{\theta}}{{\tilde{\rho}}^2} 
[(r^2 + a^2)^2 - a^2 \Delta \sin^2{\theta}] \, {\rm e}^{2 \psi} , \nonumber 
\end{eqnarray} 

\noindent
where the potentials $ V_K, \, W, \, X_K, \, Y_K, \, Z_K $ are the Lewis 
potentials for the Kerr-Newman metric, and 
$ {\tilde{\rho}}^2 = r^2 + a^2 \cos^2{\theta}. $ Also, $ J = M a$ is 
the angular momentum. 
 
\noindent
The cross term potential $ W $ is unaltered to preserve the following metric 
form

\begin{eqnarray} 
\label{superkerr}
d{s}^2 & = & - \frac{\Delta}{{\tilde{\rho}}^2} 
[{\rm e}^{- \psi} dt - a {\rm e}^{\psi} \sin^2{\theta} d \phi]^2
+ \frac{\sin^2{\theta}}{{\tilde{\rho}}^2} 
[(r^2 + a^2) {\rm e}^{\psi} d \phi - a {\rm e}^{- \psi} d t ]^2 \\
& + & {\tilde{\rho}}^2 {\rm e}^{2 \chi} \left(\frac{d r^2}{\Delta} 
+ d \theta^2 \right) .
\nonumber 
\end{eqnarray}

\noindent 
These potentials guarantee that one gets the Kerr metric if 
$ \psi = \chi = 0 $. The function $ \psi $ and $ \chi $ will be found 
approximately from the EFE.

\section{The Approximate Kerr-Newman Metric with \\ Quadrupole}
\label{sec:super}

\noindent
As was stated in the previous section our problem has been reduced to finding 
the functions $ \psi $ and $ \chi $. Here we proceed to find such functions by 
solving the EFE perturbatively. Also, we discuss the limiting cases of the new 
metric. 

\noindent 
The EFE are given by

\begin{eqnarray}
\label{einstein} 
G_{i j} & = & R_{i j} - \frac{R}{2} g_{i j} = \kappa T_{i j} , \\
\nabla_{j} F^{i j} & = & 
\frac{1}{\sqrt{- g}} \partial_j [\sqrt{- g} F^{i j}] = 0 , \nonumber
\end{eqnarray} 

\noindent
where $ G_{i j} $ ($ i, \, j = 0, \, 1,\, 2, \, 3 $) are the Einstein tensor 
components, $ R_{i j} $ are the Ricci tensor components, $ R $ is the curvature 
scalar, $ \kappa = 8 \pi G / c^4 $, $ g = {\rm det}(g_{i j}) $ and $ T_{i j} $ 
represent the energy-momentum tensor components, which are  given by

\begin{equation}
\label{stress} 
4 \pi T_{i j} = g^{k l} F_{i l} F_{j k} - \frac{1}{4} F^{a b} F_{a b} \, g_{i j} ,
\end{equation}

\noindent
where $ F_{i j} = \partial_{j} A_{i} - \partial_{i} A_{j} $ are the 
electromagnetic tensor components, and $ A_{i} $ is the vector potential 
components. 

\noindent
The $ 1 $-form for the vector potential $ {\sf A} $ can be written as follows

\begin{equation}
\label{potA}
{\sf A} = - \frac{e r}{\rho^2} \left[ {\rm e}^{- \psi} {d} {t} 
- a {\rm e}^{\psi} \sin^2{\theta} {d} {\phi} \right] .
\end{equation}

\noindent
In addition, the $ 2 $-form for the electromagnetic tensor $ {\sf F} $ can be 
obtained as follows

\begin{equation}
\label{dpotA}
{\sf F} = d {\sf A} = \frac{1}{2} F_{ij} d x^{j} \wedge d x^{j} . 
\end{equation}

\noindent
From (\ref{dpotA}) we can determine the energy-momentum tensor.

\noindent
Let us highlight the terms that are neglected in our perturbative approach, 
which are 

\begin{eqnarray}
W^2 \frac{\partial \psi}{\partial x^i} & \sim & 0, \nonumber \\ 
W \frac{\partial W}{\partial x^i} \frac{\partial \psi}{\partial x^i} 
& \sim & 0, \nonumber \\ 
W^2 \frac{\partial \chi}{\partial x^i} & \sim & 0, \nonumber \\ 
W \frac{\partial W}{\partial x^i} \frac{\partial \chi}{\partial x^i} 
& \sim & 0 . \nonumber
\end{eqnarray}
 
\noindent
Moreover, eliminating the terms corresponding to the Kerr metric into the Ricci 
tensor components, we get the Ricci tensor component of the appendix of 
\cite{Frutos2}. Note that $ W $ plays the role of the rotation, since it is proportional to the angular momentum. 
However, the above expressions do not mean that factors of $ J^2 $ or $ a^2 $ 
vanish, what is being effectively restricted here are combinations of 
quadrupoles with the angular momentum or the rotation.
 
\noindent
In order to obtain an expression for the Ricci tensor, we propose the 
following Ansatz

\begin{eqnarray}
\label{Anstaz}
\psi & = & \frac{q}{r^3} P_2 + \alpha \frac{M q}{r^4} P_2 ,\\ 
\chi & = & \frac{q P_2}{r^3} 
+ \frac{M q}{r^4} (\beta_1 + \beta_2 P_2 + \beta_3 P_2^2)
+ \frac{q^2}{r^6} (\beta_4 + \beta_5 P_2 + \beta_6 P_2^2 + \beta_7 P_2^3) , 
\nonumber
\end{eqnarray} 

\noindent
where $ q $ represents the quadrupole parameter and $ P_2 $ is the usual 
Legendre polynomial, $ P_2 = (3 \cos^2{\theta} - 1)/{2} $. Furthermore, 
the $ \alpha $ and $ \beta $'s are constants to be determined. 

\noindent
One can find the undetermined constants by substituting this Ansatz into 
the Ricci tensor components. We get a set of linear equations for these 
constants $ \alpha $, and $ \beta_n $ ($ n = 1, \, \dots, \, 7 $). After 
solving these linear equations, the constants are found to be the same as in 
\cite{Frutos4}

\begin{eqnarray}
\alpha & = & 3 \nonumber ,\\
\beta_1 & = & - \frac{1}{3} ,\nonumber \\
\beta_2 & = & \beta_3 = \frac{5}{3} ,\nonumber \\
\beta_4 & = & \frac{2}{9}, \\
\beta_5 & = & - \frac{2}{3} , \nonumber \\
\beta_6 & = & - \frac{7}{3} ,\nonumber \\
\beta_7 & = & \frac{25}{9} .\nonumber
\end{eqnarray}
 
\noindent
Finally we have all the information we require to construct our 
Kerr-Newman-like metric. From (\ref{superkerr}), the metric components are 
given by

\begin{eqnarray}
\label{metriccomp}
g_{t t} & = & \frac{{\rm e}^{- 2 \psi}}{\rho^2} [a^2 \sin^2{\theta} - \Delta], 
\nonumber \\
g_{t \phi} & = & \frac{a}{\rho^2} [\Delta - (r^2 + a^2)] \sin^2{\theta} 
= \frac{\sin^2{\theta}}{\rho^2} (aQ^2 - 2Jr),\\
g_{r r} & = & \rho^2 \frac{{\rm e}^{2 \chi}}{\Delta}, \nonumber \\
g_{\theta \theta} & = & \rho^2 {\rm e}^{2 \chi} ,\nonumber \\
g_{\phi \phi} & = & \frac{{\rm e}^{2 \psi}}{\rho^2} 
[(r^2 + a^2)^2 - a^2 \Delta \sin^2{\theta}] \sin^2{\theta} ,
\nonumber 
\end{eqnarray} 

\noindent 
We checked that (\ref{metriccomp}) was valid up to order 
$ O(a q^2, \, a^2 q, \, M a q, \, M q^2, \, M^2 q, \, q^3) $ using a REDUCE 
program \cite{Hearn}. 

\noindent 
Now, we will focus in the limiting cases of the new metric. We summarized the 
limiting cases in table \ref{tab:table}. First, note that if $ e = 0 $ 
in (\ref{metriccomp}) one recovers the metric found by Frutos {\it et al.} 
in \cite{Frutos2}. Therefore, following \cite{Frutos2} the other interesting 
limiting cases are: The Kerr metric if $ e = q = 0 $, the metric found in 
\cite{Frutos1} if $ e = a^2 = q^2 = 0 $, the Erez-Rosen-like metric described 
in \cite{Frutos2} if $ a = e = 0 $, and the Schwarzchild metric if 
$ a = e = q = 0 $. Furthermore, one obtains the Kerr-Newman geometry if 
$ q = 0 $. Also, the Reissner-Nordstr\"om metric is obtained if $a = q = 0 $. 
Thus, all the expected limiting cases can be obtained from this new metric.

\begin{table}[t!]
\centering
\begin{tabular}{l|c|l}
\hline \hline
Absent physical property & Small physical property & Limiting metric\\
\hline \hline
Charge  & Quadrupole (linear) & Metric found in  \cite{Frutos2} \\
\hline \hline
Charge  & Quadrupole (quadratic) & Metric found in \cite{Frutos4} \\
\hline
Quadrupole & - & Kerr-Newman \\
\hline
Quadrupole and rotation  & - & Reissner-Nordstr\"om \\
\hline
Charge and quadrupole & - & Kerr \\
\hline
Charge and rotation & - & Erez-Rosen-like \\
\hline
Charge & Quadrupole and rotation & Metric found in \cite{Frutos1} \\
\hline
Charge, quadrupole and rotation & - & Schwarzschild \\
\hline \hline
\end{tabular}
\caption{Limiting cases.}
\label{tab:table}
\end{table}

\noindent 
Here we will not show the matching of (\ref{metriccomp}) with the HT metric. 
This matching is vital because it guarantees that an interior solution exists. 
We will not show it since it should be trivial to follow \cite{Frutos2} since 
our metric has the same form, or equivalently follow \cite{Frutos3}. Moreover, 
in \cite{Frutos4}, it was shown that the multipole structure of this metric 
without charge is non-isometric with the Quevedo-Mashhoon \cite{QM,Quevedo} 
and the Manko-Novikov \cite{Manko} metrics. Then, this new metric should not 
be isometric with charged version of these metrics.

\section{Conclusion}
\label{sec:conclusion}

\noindent 
A metric with charge, deformations and rotation was obtained by solving the 
EFE pertrubatively. The expected limiting cases of this new metric were 
explored and found, which is a positive sign. Moreover, these limiting cases 
confirm that our metric adequately describes a mass with charge and quadrupole 
under rotation. We successfully applied the perturbation method developed by 
Frutos et al in \cite{Frutos1,Frutos2,Montero,Frutos4} to obtain a new 
approximate metric. Notice that the main improvement of our work with respect 
to \cite{Frutos2,Frutos4} is the inclusion of charge. Ideally, this moves us 
closer to representing an actual astrophysical object. Further, one could 
expect that careful understanding of this procedure might eventually lead us 
towards a magnetized object.  

\noindent 
Even though we did not explicitly showed that the new metric can be matched to 
the Hartle-Thorne metric the previous success in the the non-charged metrics 
\cite{Frutos2,Frutos4} is encouraging. Particularly, because of the similarity 
with that non-charged metric we do not expect any issues in the matching. This 
matching is crucial for the sake of showing that an interior metric for our 
new metric exists. 

\noindent 
Since this new metric can describe real charged astrophysical objects in a more 
realistic fashion than the Kerr-Newman or Reissner-Nordstr\"om metrics, 
then our metric should be attractive for astrophysical applications, like 
gravitational lensing and relativistic magnetohydrodynamics. Furthermore, 
computational implementation of this metric should not imply additional 
difficulties since it mantains a similar form to the Kerr metric.  


\end{document}